\documentclass[sigconf]{acmart}
\usepackage{multirow}

\copyrightyear{2026}
\acmYear{2026}
\setcopyright{cc}
\setcctype{by}
\acmConference[ICMI Companion '26]{Companion of the INTERNATIONAL CONFERENCE ON MULTIMODAL INTERACTION}{October 05--09, 2026}{Napoli, Italy}
\acmBooktitle{Companion of the INTERNATIONAL CONFERENCE ON MULTIMODAL INTERACTION (ICMI Companion '26), October 05--09, 2026, Napoli, Italy}
\acmDOI{10.1145/3776591.3837045}
\acmISBN{979-8-4007-2319-3/2026/10}

\begin{document}

\title{Multimodal Temporal Modeling for Continuous Group Emotion Recognition in Multi-party Dialogues}

\author{Soma Iwata}
\orcid{0009-0006-2452-8004}
\affiliation{%
  \institution{Kyoto University}
  \city{Kyoto}
  \country{Japan}
}
\email{iwata@sap.ist.i.kyoto-u.ac.jp}

\author{Koji Inoue}
\orcid{0000-0002-2929-2559}
\affiliation{%
  \institution{Kyoto University}
  \city{Kyoto}
  \country{Japan}
}
\email{inoue@sap.ist.i.kyoto-u.ac.jp}

\author{Muyun Wu}
\orcid{0009-0003-4224-5452}
\affiliation{%
  \institution{Kyoto University}
  \city{Kyoto}
  \country{Japan}
}
\email{wu@sap.ist.i.kyoto-u.ac.jp}

\author{Taiga Mori}
\orcid{0000-0001-6993-5726}
\affiliation{%
  \institution{Kyoto University}
  \city{Kyoto}
  \country{Japan}
}
\email{mori@sap.ist.i.kyoto-u.ac.jp}

\author{Divesh Lala}
\orcid{0009-0005-3181-6410}
\affiliation{%
  \institution{Kyoto University}
  \city{Kyoto}
  \country{Japan}
}
\email{lala@sap.ist.i.kyoto-u.ac.jp}

\author{Tatsuya Kawahara}
\orcid{0000-0002-2686-2296}
\affiliation{%
  \institution{Kyoto University}
  \city{Kyoto}
  \country{Japan}
}
\email{kawahara@i.kyoto-u.ac.jp}

\renewcommand{\shortauthors}{Iwata et al.}

\begin{abstract}

To realize natural behavior in dialogue agents in multi-party dialogue scenarios, it is important to understand group emotion such as valence and arousal as a whole.
Most prior work addressed this task at the utterance level or using a coarse-grained time window, which is not sufficient to capture emotional dynamics.
In this study, we formulate continuous recognition of the Group Emotion at a one-second resolution.
Moreover, we also introduce the Mixed state, which captures the emotional divergence among participants in the group.
We constructed a dataset with frame-level soft labels based on the TEIDAN corpus and propose a multimodal temporal framework that integrates audio and video information using a sliding-window context.
Experimental results demonstrate that the temporal Transformer outperforms simple baselines and shows stronger temporal agreement with the ground-truth labels than the LLM-based model.
The effect of context length is limited, whereas audio-visual input outperforms either unimodal input on the continuous-label metrics.
Additionally, our analysis shows larger Group Emotion recognition errors in intervals with high Mixed values, exposing emotional divergence as a key challenge for group emotion recognition.

\end{abstract}



\maketitle

\section{Introduction}\label{sec-intro}

To enable dialogue agents to interact naturally with humans, it is crucial to understand not only linguistic information but also the emotions and intentions of the dialogue partners, as well as the dialogue atmosphere \cite{irfan2020dynamic}.
In dyadic dialogues, emotions can generally be recognized from a single user's facial expressions, tone of voice, and utterance content.
In multi-party dialogues, however, interactions involve multiple human participants alongside the agent, who mutually influence one another \cite{cooney2020many}.
Consequently, the emotional state of the group as a whole is not merely the sum of each participant's individual emotions; rather, it emerges from the interactions among the participants \cite{barsade2015,park2015group}.
Furthermore, emotions can change within an utterance and during periods in which no participant is speaking.
Especially in multi-party scenarios, where interruptions occur and multiple listeners are present, emotional dynamics become temporally more complex than in dyadic dialogues \cite{kumar2023emotion}.
Therefore, for an agent to behave appropriately in multi-party dialogues, it is essential to continuously recognize the group emotion formed through these interactions.

According to Barsade and Knight \cite{barsade2015}, there are four main components of group emotion: convergence, divergence, emotional culture, and dynamic process.
Convergence refers to the sharing of the same emotion among group members, while divergence refers to the presence of differing emotions among them.
These are bottom-up aspects determined by individual emotions.
Emotional culture consists of implicit rules regarding which emotions should be expressed or suppressed within a group, representing a top-down aspect determined by the group's context.
Finally, dynamic process refers to how group emotions change over time.
This aspect exhibits both top-down characteristics, as the group emotion guides the direction of individual emotions, and bottom-up characteristics, as individual emotions collectively constitute the group emotion.
Considering these elements is vital for modeling natural interactions.

\begin{table*}[t]
  \centering
  \caption{Comparison of problem settings between existing studies and our study}
  \label{tab:related-work-comparison}
  \small
  \begin{tabular}{lccccc}
    \toprule
    Study & Scale & Target & Temporal Resolution & Consideration of Non-speakers & Divergence \\
    \midrule
    IEMOCAP \cite{busso2008} & Dyadic & Individual & Turn & $\times$ & $\times$ \\
    MELD \cite{poria2019} & \textbf{Multi-party} & Individual & Utterance & $\times$ & $\times$ \\
    MEmoR \cite{shen2020} & \textbf{Multi-party} & Individual & Clip & \textbf{\checkmark} & $\times$ \\
    RECOLA \cite{ringeval2013} & Dyadic & Individual & \textbf{Continuous} & \textbf{\checkmark} & $\times$ \\
    VGAF-GEMS \cite{kataria2025gems} & \textbf{Multi-party} & \textbf{Multi-level} & \textbf{Frame} / 5s clip & \textbf{\checkmark} & $\times$ \\
    GCE \cite{lim2024gce} & \textbf{Multi-party} & \textbf{Group} & 30s clip & \textbf{\checkmark} & $\times$ \\
    Prabhu et al. \cite{prabhu2025} & \textbf{Multi-party} & \textbf{Group} & 15s window & \textbf{\checkmark} & \textbf{\checkmark} \\
    \midrule
    \textbf{Ours} & \textbf{Multi-party} & \textbf{Group} & \textbf{1s} & \textbf{\checkmark} & \textbf{\checkmark} \\
    \bottomrule
  \end{tabular}
\end{table*}

Table \ref{tab:related-work-comparison} summarizes the comparison between existing studies and our work based on the participant scale (dyadic/multi-party), recognition target, temporal resolution, consideration of non-speakers, and explicit treatment of emotional divergence.
IEMOCAP \cite{busso2008}, a representative dataset for emotion recognition in conversation (ERC), mainly handles the individual emotion of the speaker at the turn level.
MELD \cite{poria2019} and EmotionLines \cite{hsu2018emotionlines} include multi-party dialogues, but their primary focus also remains on the individual emotion of the speaker at the utterance level.
MEmoR \cite{shen2020} extends individual emotion recognition to listeners in multi-party dialogues, but still predicts emotion at the individual level.
RECOLA \cite{ringeval2013} instead captures continuous emotional changes, but focuses on individual emotion in dyadic interactions.
Together, these studies broaden individual emotion recognition in terms of both participants and temporal resolution, but do not directly model the collective emotional state of a multi-party group.

Recent studies have begun to address emotion directly at the group level.
VGAF \cite{sharma2023vgaf} comprises 4,183 approximately five-second video clips with positive, neutral, and negative group-affect labels.
Building on VGAF, GEMS \cite{kataria2025gems} adds frame-level emotion annotations for each group member and models emotion at the individual, group, and event levels.
Its videos cover diverse social situations rather than focusing specifically on dialogue.
GCE \cite{lim2024gce} provides group emotion and cohesion annotations for 30-second audio-visual conversational clips.
Both datasets treat each video clip as an independent sample rather than tracking group emotion throughout an ongoing dialogue.

Moving from clip-level group understanding toward temporal dialogue modeling, Prabhu et al. \cite{prabhu2025} analyzed dynamics such as emotional convergence and divergence among group members.
Their 15-second windows capture longer-term group dynamics, but are too coarse to identify when emotions change within an utterance or between turns.
To the best of our knowledge, no prior work combines recognition of Group Emotion at one-second intervals over the full course of multi-party dialogues with explicit annotation and analysis of emotional divergence.

In this study, we address continuous Group Emotion recognition over the full course of multi-party dialogues.
We represent Group Emotion at each time point along two dimensions, Valence and Arousal, and introduce Mixed Arousal and Mixed Valence to annotate emotional divergence among participants.
We continuously annotate 36 dialogues from the TEIDAN corpus and aggregate the annotations into one-second soft labels.
This temporal granularity allows emotional changes to be aligned with nearby utterances and nonverbal events, providing a basis for investigating their potential triggers.
Using this dataset, we compare a temporal Transformer with an LLM-based model that integrates participant-specific audio and visual features over a causal sliding-window context.
We compare five- and 20-second context windows and audio-visual, audio-only, and video-only conditions to investigate the effects of context length and input modality.
We further analyze recognition performance according to the Mixed annotations to examine how emotional divergence relates to the difficulty of Group Emotion recognition.

\section{Dataset}

This section describes the TEIDAN corpus used in this study, the annotation design, and the procedure for constructing the ground-truth labels used for model input.
Furthermore, we analyze the label distribution and inter-annotator agreement, and discuss the relationship between dialogue context and emotion labels.

\subsection{TEIDAN Corpus}\label{subsec-corpus}
\begin{figure}[tb]
  \centering
  \includegraphics[width=\linewidth]{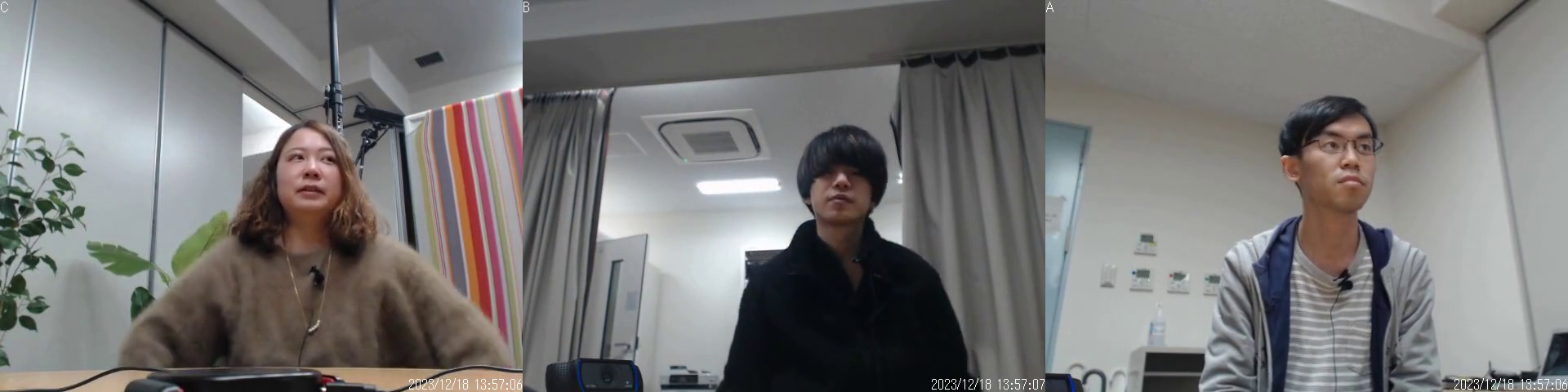}
  \caption{Example of video data from the TEIDAN corpus. A split-screen view shows the frontal views of all three participants.}
  \Description{A split-screen video frame shows three frontal views of the participants during the same dialogue session.}
  \label{fig-corpus-example}
\end{figure}

In this study, we utilized the Japanese portion of the TEIDAN corpus \cite{mori2026teidan}, a multilingual multi-party dialogue dataset comprising recordings of triadic interactions.
This corpus consists of interactions among groups of three members from the same laboratory, primarily consisting of undergraduate and graduate students, discussing given topics.
Each group performed either spontaneous discussions or attentive listening for three different topics.
For our study, we used data from all 12 groups who participated in the spontaneous discussions, resulting in a total of 36 dialogues that lasted approximately six minutes on average.
The dataset includes frontal video recordings of each participant, as shown in Figure \ref{fig-corpus-example}, and audio data captured via individual lavalier microphones.

\subsection{Annotation Design}\label{subsec-annotation-design}
We conducted group emotion annotation for the 36 spontaneous discussions.
Specifically, five annotators assigned four types of labels to each dialogue: Arousal, Valence, Mixed Arousal, and Mixed Valence.
Here, Arousal and Valence represent the emotional state of the group (hereafter referred to as \textit{Group Emotion labels}).
To capture the element of divergence, Mixed Arousal and Mixed Valence are assigned to segments where the participants' emotions diverge substantially from one another (hereafter referred to as \textit{Mixed labels}).

While many emotion recognition studies employ utterance-level annotation, this study adopts time-continuous annotation without fixed units.
This approach is motivated by two main reasons.
First, it allows us to capture emotional changes that occur during a single utterance.
In long utterances, the group's emotional state may shift midway; continuous tracking is essential to understand such transitions.
Second, it enables the capture of emotions during silences.
Silences can range from an awkward atmosphere where no one wants to speak to a relaxed one after a topic concludes.
Since utterance-level annotation typically excludes silent segments, annotation without fixed units is necessary to capture these nuances.
Furthermore, continuous annotation addresses the dynamic process of group emotion mentioned in Section \ref{sec-intro}.
This annotation design allows a model to use a long dialogue context as input while predicting Group Emotion at a fine temporal resolution.

The criteria for the four types of annotations are as follows.
For Group Emotion labels, we extended Russell's circumplex model \cite{russell1980} from an individual to a group scale, representing the emotional state of the group through Arousal and Valence.
Each was classified into five levels from +2 to -2.
The original label definitions were provided in Japanese; their English translations are shown in Tables \ref{tab-arousal-definition} and \ref{tab-valence-definition}, which are consistent with the LLM prompts used later in this study.
The Mixed label is assigned to segments where the emotions of the participants diverge.
The definitions of Mixed labels are shown in Table \ref{tab-mixed-definition}.

\begin{table}[tbp]
\centering
\caption{Label definitions for Arousal}
\label{tab-arousal-definition}
\begin{tabular}{cp{0.78\linewidth}}
\toprule
Class & Definition \\
\midrule
+2 & The group is extremely energetic.\\
+1 & The emotional energy of the group is high.\\
0 & Calm and stable state.\\
-1 & The energy level of the group is low.\\
-2 & The group activity is almost stopped.\\
\bottomrule
\end{tabular}
\end{table}

\begin{table}[tbp]
\centering
\caption{Label definitions for Valence}
\label{tab-valence-definition}
\begin{tabular}{cp{0.78\linewidth}}
\toprule
Class & Definition \\
\midrule
+2 & Strong and clear positive emotion shared by the group.\\
+1 & Positive atmosphere flowing in the group.\\
0 & No strong emotional color observed.\\
-1 & Slight negative emotion drifting in the group.\\
-2 & The group is dominated by strong discomfort.\\
\bottomrule
\end{tabular}
\end{table}

\begin{table}[tbp]
\centering
\caption{Definitions of Mixed labels}
\label{tab-mixed-definition}
\setlength{\tabcolsep}{2pt}
\begin{tabular}{p{0.25\columnwidth}p{0.70\columnwidth}}
\toprule
Class & Definition \\
\midrule
Mixed Arousal & Arousal levels differ among participants, and a single group-level Arousal cannot be identified.\\
\midrule
Mixed Valence & Positive and negative emotions coexist, and a single group-level Valence cannot be identified.\\
\bottomrule
\end{tabular}
\end{table}

\begin{figure}[tb]
  \centering
  \includegraphics[width=\linewidth]{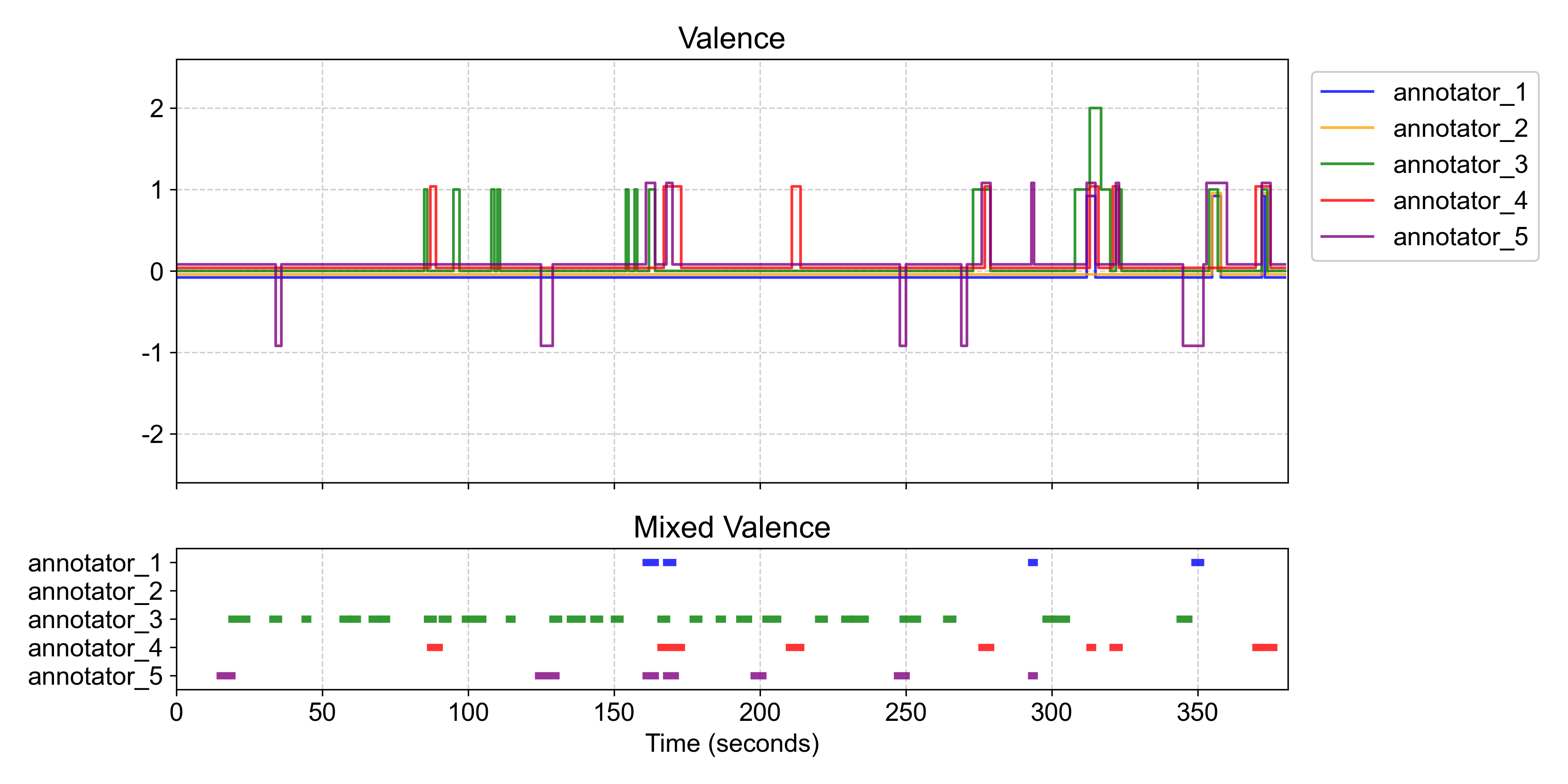}
  \caption{Example of continuous annotation for Valence and Mixed Valence by five annotators.}
  \Description{Two stacked plots share an x-axis labeled Time in seconds from 0 to about 350. The top plot, titled Valence, shows five colored step lines for annotators 1 through 5, taking discrete values between minus 1 and plus 2, remaining near 0 for most of the session with occasional positive spikes. The bottom plot, titled Mixed Valence, shows short horizontal bars for each annotator indicating time spans labeled as mixed; annotator 3 has many segments across the session while the others have fewer.}
  \label{fig-annotation-timeseries}
\end{figure}

During the task, annotators were presented with split-screen videos and mixed audio from all participants.
They were instructed to make holistic judgments using all observable information.
Importantly, they were restricted to using information only up to the current moment, without considering future context.
Even in segments labeled as Mixed, annotators were required to assign Arousal and Valence labels ranging from -2 to +2.
Because there were no fixed units like utterances, annotators freely determined the start and end points of any segments they judged to be non-neutral or mixed.
Unmarked periods were treated as 0 or Not Mixed.
Figure \ref{fig-annotation-timeseries} shows an example of the five annotators' time-series data for Valence and Mixed Valence in a single dialogue.
The upper plot shows Valence classes (-2 to +2), and the lower bars indicate segments identified as Mixed by each annotator.
While each annotator continuously assigns labels based on their individual judgment, the plot demonstrates a reasonable agreement among them on the timing of the Valence peaks.

\subsection{Label Construction}\label{subsec-label-construction}
This section describes the procedure for creating ground-truth labels for model training and evaluation from the continuous annotation data.

\subsubsection{Data Selection}\label{subsec-annotator-selection}
Regarding the Mixed labels, to ensure annotation quality, we excluded annotations whose total labeled durations were outliers among the five annotators.
As a criterion, an annotator's data was discarded if the difference between their assigned time and the mean assigned time of the five annotators was $1.5\sigma$ or greater (where $\sigma$ is the standard deviation of the total assigned time across the five annotators).
This process was performed independently for Mixed Arousal and Mixed Valence for each dialogue, and excluded one annotator in all 36 dialogues for Mixed Arousal and in 27 dialogues for Mixed Valence.
For example, in the Mixed Valence plot of Figure \ref{fig-annotation-timeseries}, Annotator 3's labeled duration is notably longer than the others.
Therefore, Annotator 3's data for this dialogue would be excluded.

\subsubsection{Aggregation into Frames}\label{subsec-data-preparation}
Using the selected data, we constructed ground-truth labels for each 1-second frame.
For a frame $t$ (length $T=1.0$s), let $\tau_{t,i,c}$ be the total duration during which annotator $i$ assigned class $c$ ($c \in \{-2, \dots, +2\}$).
Since Group Emotion labels utilize all five annotators, we set $N=5$.
The soft label $y_{c}^{(t)}$ for class $c$ in frame $t$ is defined as the average time occupancy across all annotators as
\begin{equation}\label{eq-soft-label}
y_{c}^{(t)} = \frac{1}{N} \sum_{i=1}^{N} \frac{\tau_{t,i,c}}{T} \, .
\end{equation}
In continuous annotation, class transitions often occur within a single frame, and boundaries may vary between annotators.
We use soft labels as training targets to preserve these fluctuations.
As described in Section \ref{subsec-experiment-setup}, our evaluation uses both regression-based and classification-based metrics.
The hard label $y_{hard}^{(t)}$ used for classification is determined by taking the argmax of the soft labels.
Mixed labels from the selected annotators were similarly aggregated into 1-second frames.
This results in continuous values in the range [0, 1], representing the time occupancy of the Mixed state within the frame.

\subsection{Label Distribution and Agreement}\label{subsec-statistics-agreement}
This section reports the label distribution and inter-annotator agreement after the processing described in Section \ref{subsec-label-construction}.

\subsubsection{Label Distribution}\label{subsubsec-label-distribution}
Before constructing model windows, we split the 12 groups into Train, Val, and Test sets in an 8:2:2 ratio.
All three topic dialogues for each group were assigned to the same split, resulting in 24/6/6 dialogues and 8,908/2,247/2,240 one-second frames for Train/Val/Test, respectively.
Consequently, no frames, dialogues, or three-person groups are shared across splits.
Although all dialogues from each group were assigned to the same split, some participants appeared in multiple splits because each participant took part in two groups.
For each split, we report the distribution of the five Arousal/Valence classes and the Mixed Arousal/Valence states, computed by summing $\tau_{t,i,c}$ (Section \ref{subsec-data-preparation}) over all frames and annotators and normalizing.

Tables \ref{tab-label-distribution-group-emotion} and \ref{tab-label-distribution-mixed} show their distributions.
For Group Emotion labels, Class 0 accounts for more than 85\%, while for Mixed labels, the Not Mixed state accounts for over 92\%.
Furthermore, Classes -2 and -1 are rarely present in the Group Emotion labels.
Our strategies for addressing this class imbalance are detailed in Section \ref{subsec-imbalance-mitigation}.
The label proportions are broadly similar across the dataset splits.

\begin{table}[tbp]
  \centering
  \caption{Label distributions for the Group Emotion labels.}
  \label{tab-label-distribution-group-emotion}
  \setlength{\tabcolsep}{2.3pt}
  \begin{tabular}{@{}crrrccrrr@{}}
  \toprule
  \multicolumn{4}{c}{\textbf{(a) Arousal [\%]}} & & \multicolumn{4}{c}{\textbf{(b) Valence [\%]}} \\
  \cmidrule(r){1-4} \cmidrule(l){6-9}
  Class & Train & Val & Test & & Class & Train & Val & Test \\
  \midrule
  -2 & 0.0 & 0.1 & 0.0 & & -2 & 0.0 & 0.0 & 0.0 \\
  -1 & 1.3 & 1.0 & 1.4 & & -1 & 0.7 & 0.2 & 0.1 \\
   0 & 88.8 & 89.1 & 89.4 & &  0 & 85.4 & 85.4 & 85.9 \\
  +1 & 8.5 & 8.6 & 7.7 & & +1 & 12.0 & 11.1 & 10.2 \\
  +2 & 1.4 & 1.3 & 1.6 & & +2 & 1.9 & 3.3 & 3.7 \\
  \bottomrule
  \end{tabular}
\end{table}

\begin{table}[tbp]
  \centering
  \caption{Label distributions for the Mixed labels.}
  \label{tab-label-distribution-mixed}
  \setlength{\tabcolsep}{2.3pt}
  \begin{tabular}{@{}lrrrclrrr@{}}
  \toprule
  \multicolumn{4}{c}{\textbf{(a) Mixed Arousal [\%]}} & & \multicolumn{4}{c}{\textbf{(b) Mixed Valence [\%]}} \\
  \cmidrule(r){1-4} \cmidrule(l){6-9}
  Class & Train & Val & Test & & Class & Train & Val & Test \\
  \midrule
  Not Mixed & 94.2 & 95.3 & 94.5 & & Not Mixed & 93.9 & 92.5 & 94.2 \\
  Mixed & 5.8 & 4.7 & 5.5 & & Mixed & 6.1 & 7.5 & 5.8 \\
  \bottomrule
  \end{tabular}
\end{table}

\subsubsection{Inter-annotator Agreement}\label{subsubsec-inter-annotator-agreement}
We evaluated the inter-annotator agreement across the entire dataset using Krippendorff's $\alpha$ \cite{hayes2007}.
Table \ref{tab-inter-annotator-agreement} shows the $\alpha$ values for each label before and after the data selection.
For Group Emotion, although the $\alpha$ values appear relatively modest due to natural temporal misalignments inherent to continuous annotation, the timing of the emotional peaks was broadly consistent among the annotators, as illustrated in Figure \ref{fig-annotation-timeseries}.
Furthermore, the scores for the Mixed labels highlight the inherently subjective challenge of recognizing emotional divergence among participants.
This selection increased agreement for both Mixed labels, although agreement remained low for Mixed Valence.

\begin{table}[tbp]
\centering
\caption{Krippendorff's $\alpha$ before and after data selection}
\label{tab-inter-annotator-agreement}

\begin{tabular}{lrr}
\toprule
Task & {$\alpha$ (Before)} & {$\alpha$ (After)} \\
\midrule
Arousal & 0.2975 & 0.2975 \\
Valence & 0.3561 & 0.3561 \\
Mixed Arousal & $-0.0505$ & 0.2650 \\
Mixed Valence & 0.0722 & 0.1011 \\
\bottomrule
\end{tabular}
\end{table}

\subsection{Relationship between Dialogue Context and Emotion Labels}\label{subsec-label-dialog-correlation}
This section characterizes what observable dialogue situations are reflected in the newly collected continuous group-level labels.

\subsubsection{Laughter and Emotion Labels}
First, we discuss the relationship between laughter and the annotations.
Laughter segments were obtained from the annotations originally included in the TEIDAN corpus.
Table \ref{tab:laughter-emotion-correlation} shows the relationship between the number of simultaneously laughing participants (0 to 3) and the average values of each emotion label.
Focusing on the Group Emotion labels, both Arousal and Valence increase as the number of laughing participants increases.
We used linear regression to examine the associations between the number of simultaneously laughing participants and Arousal and Valence.
To account for dependencies among one-second frames and the three dialogues recorded from each group, we resampled the 12 three-person groups with replacement 10,000 times and refitted each regression to every bootstrap sample to obtain percentile 95\% CIs.
Each additional laughing participant was associated with an increase of 0.245 in Arousal (95\% CI $[0.206, 0.277]$) and 0.323 in Valence ($[0.278, 0.363]$).

Regarding the Mixed labels, Mixed Arousal peaked at 0.32 when two people laughed, and decreased to 0.18 when three people laughed.
Similarly, Mixed Valence maintained relatively high values of 0.16 when only one or two people laughed, but dropped to 0.06 when all three laughed simultaneously.
These descriptive results suggest greater emotional divergence when only one or two participants laugh than when all three laugh, although simultaneous laughter by two or three participants was relatively rare.

\begin{table}[tb]
  \centering
  \caption{Comparison of average values of each emotion label by the number of simultaneously laughing participants}
  \label{tab:laughter-emotion-correlation}

  \footnotesize
  \setlength{\tabcolsep}{4.5pt}
  \begin{tabular}{crrrrr}
\toprule
    Count & {Duration [s]} & {Arousal} & {Valence} & {Mixed Arousal} & {Mixed Valence} \\
\midrule
    0 & 11,024 & 0.04 & 0.08 & 0.02 & 0.04 \\
    1 & 1,531  & 0.21 & 0.34 & 0.20 & 0.16 \\
    2 & 618   & 0.57 & 0.77 & 0.32 & 0.16 \\
    3 & 222   & 0.86 & 1.11 & 0.18 & 0.06 \\
\bottomrule
  \end{tabular}
\end{table}

\subsubsection{Dialogue Activity and Emotion Labels}
Next, we investigate how dialogue activity, such as dialogue pace and silence, is associated with emotion labels by analyzing turn-initiation frequency and silence.

\paragraph{Analysis Conditions and Metric Definitions}
The definitions for turn-initiation frequency and silence are as follows:

\begin{enumerate}
  \item \textbf{Turn-initiation frequency}: \\
  This metric quantifies the pace of dialogue and the speed of speaker transitions.
  For frame-level analysis, the interval from time $t$ to $t+1$ is defined as frame $t$, and the frame average is calculated as the average occupancy of the label.
  We define an analysis window $W_t$ spanning the preceding 3 seconds $[t-2, t+1]$ for a frame $t$ ending at $t+1$.
  Based on the Turn annotations in the TEIDAN corpus, a \textit{turn-initiation} event is defined as the point when a turn by a different participant starts.
  The total number of turn-initiation events within $W_t$ is defined as the turn-initiation frequency for frame $t$.
  Note that since turns can overlap among participants, a turn-initiation does not necessarily perfectly align with turn-taking.

  \item \textbf{Silence}: \\
  A \textit{silence segment} is defined as a continuous period of at least one second where none of the three participants are speaking.
  All other periods are defined as non-silence segments.
  For segment-level analysis, the average values are calculated as the average occupancy of each label over the entire duration of the respective segment.
\end{enumerate}

\begin{table}[tb]
  \centering
  \caption{Comparison of average values of each emotion label by turn-initiation frequency}
  \label{tab:turn-initiation-metrics}

  \footnotesize
  \setlength{\tabcolsep}{4.5pt}
  \begin{tabular}{crrrrr}
\toprule
    Count & {Duration [s]} & {Arousal} & {Valence} & {Mixed Arousal} & {Mixed Valence} \\
\midrule
    0 & 7,454 & 0.054 & 0.111 & 0.045 & 0.054 \\
    1 & 3,993 & 0.131 & 0.198 & 0.067 & 0.072 \\
    2 & 1,553 & 0.196 & 0.264 & 0.072 & 0.075 \\
    3 & 348  & 0.246 & 0.312 & 0.080 & 0.088 \\
    4 & 44   & 0.249 & 0.250 & 0.061 & 0.070 \\
    5 & 3    & 0.133 & 0.338 & 0.250 & 0.172 \\
\bottomrule
  \end{tabular}
\end{table}

\begin{table}[tb]
  \centering
  \caption{Comparison of average values of each emotion label by presence of silence}
  \label{tab:silence-analysis}
  \footnotesize
  \setlength{\tabcolsep}{3pt}
  \begin{tabular}{lrrrrr}
\toprule
    State & {Duration [s]} & {Arousal} & {Valence} & {Mixed Arousal} & {Mixed Valence} \\
\midrule
    Non-silence & 12,816.8 & 0.107 & 0.169 & 0.058 & 0.065 \\
    Silence    & 578.2   & $-0.077$ & $-0.026$ & 0.003 & 0.022 \\
\bottomrule
  \end{tabular}
\end{table}
Table \ref{tab:turn-initiation-metrics} shows the relationship between turn-initiation frequency and emotion labels.
Excluding the sparsely represented counts of 4 and 5 from the category-wise comparison, both Arousal (0.054 $\to$ 0.246) and Valence (0.111 $\to$ 0.312) increased monotonically as the turn-initiation frequency increased from 0 to 3.
We applied the same regression and group-level cluster bootstrap procedure to turn-initiation count.
Each additional turn initiation was associated with an increase of 0.068 in Arousal (95\% CI $[0.056, 0.079]$) and 0.073 in Valence ($[0.056, 0.092]$).

Table \ref{tab:silence-analysis} shows the relationship between silence and emotion labels.
Focusing on silence segments, Group Emotion labels show negative values.
While silences can occur for various reasons, the Group Emotion labels during these periods tended to be negative on average in this dataset.
A group-level cluster bootstrap showed that silence had lower Arousal (difference $=-0.184$, 95\% CI $[-0.218, -0.145]$) and Valence (difference $=-0.195$, $[-0.230, -0.156]$) than non-silence.

\section{Proposed Model}

This section describes the models used for continuous Group Emotion recognition in multi-party dialogues.

\subsection{Model Architecture}\label{subsec-model-structure}

\begin{figure}[tbp]
    \centering
    \includegraphics[width=1.0\columnwidth]{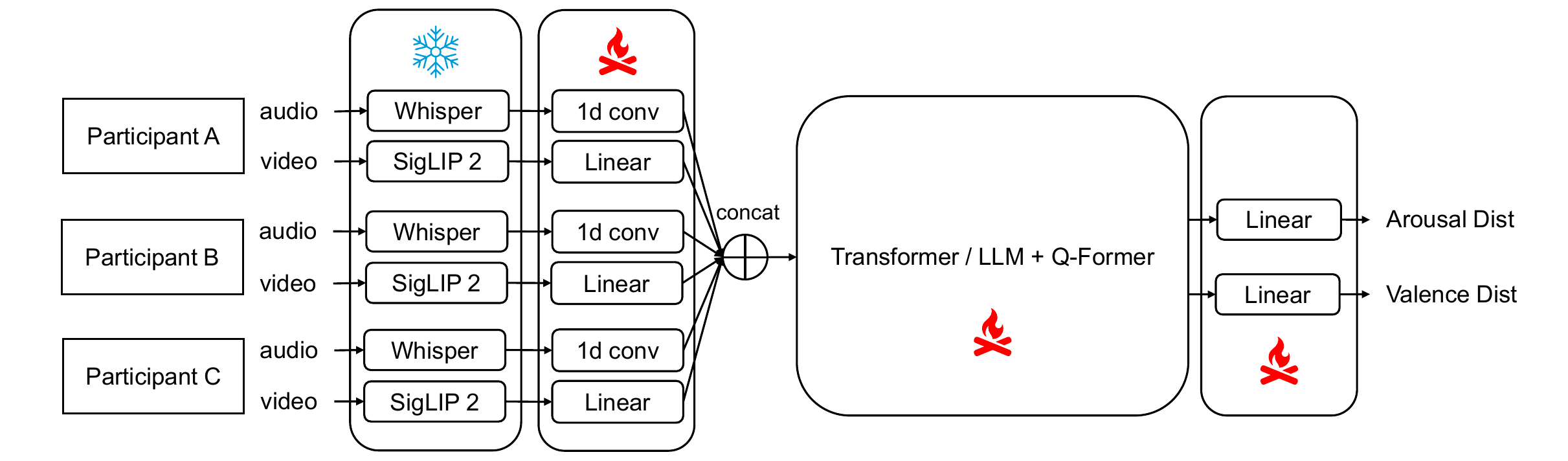}
    \caption{Overview of the model architectures.}
    \Description{For each of three participants, audio and video are processed by frozen Whisper and SigLIP 2 encoders, followed by trainable 1D convolution and linear projection layers. The participant-specific features are concatenated and processed by either a temporal Transformer or an LLM with a Q-Former. Two linear heads output the predicted Arousal and Valence distributions.}
    \label{fig-architecture}
\end{figure}

Figure \ref{fig-architecture} presents an overview of the two model variants evaluated in this study: a temporal Transformer and an LLM-based model.
Both variants share the same participant-specific audio and visual feature extraction pipeline.
First, we extract visual features using SigLIP 2 \cite{tschannen2025} and audio features using the Whisper encoder \cite{radford2023whisper} separately for each participant.
Since SigLIP 2 is an image encoder, we extract images from the video frame by frame and apply SigLIP 2 to each frame.
For each frame, the visual feature is obtained as a 768-dimensional vector, and the audio feature as a 1024-dimensional vector.
The temporal resolution of feature extraction is 30Hz for video and 50Hz for audio.
Because the output of SigLIP 2 consists of patch-level features, we calculate the average across all patches for each frame and each participant, followed by a linear layer.
This linear layer shares weights across participants, and the dimension remains 768 after its application.

For the audio features, a 1D convolution with a kernel size of 2 and a stride of 2 is applied to downsample the temporal resolution from 50Hz to 25Hz.
This 1D convolution also shares weights across participants.
Subsequently, the audio features are upsampled to 30Hz via linear interpolation to temporally align them with the visual features.
The resulting six sequences (visual and audio modalities for each of the three participants) at 30Hz are concatenated frame by frame to obtain a $3 \times (768 + 1024) = 5376$-dimensional multimodal feature.

In the temporal Transformer, the concatenated feature is projected to 768 dimensions and combined with sinusoidal positional encodings.
The resulting sequence is processed by six Transformer encoder layers, each with 12 attention heads and a 3072-dimensional feed-forward layer, followed by layer normalization.
We average the Transformer outputs over the final one-second interval and apply separate linear layers with softmax functions to predict the Arousal and Valence distributions.

For comparison, we also evaluate an LLM-based model that extends SilenceLLM proposed by Wu et al. \cite{wu2026} to support multi-party dialogues and continuous Group Emotion recognition.
The concatenated multimodal feature is fed into a Q-Former \cite{li2023}, which uses learnable query vectors to extract information through cross-attention.
The Q-Former output is projected to the embedding dimension of Qwen3-1.7B \cite{yang2025qwen3} and provided to the LLM together with an instruction prompt.
Two linear layers with softmax functions are applied to the final hidden representation to predict the Arousal and Valence distributions.
The prompt is provided in the supporting file.

Regarding the training strategy, the parameters of SigLIP 2 and the Whisper encoder are frozen and not updated during training.
The modality projection layers and output heads are trained from scratch in both variants, as are all layers of the temporal Transformer.
For the LLM-based model, the Q-Former and its projection layer are also trained from scratch, while the query, key, value, and output projection matrices in Qwen3-1.7B are adapted using LoRA \cite{hu2022}.

\begin{figure}[tbp]
	\centering
	\includegraphics[width=\columnwidth]{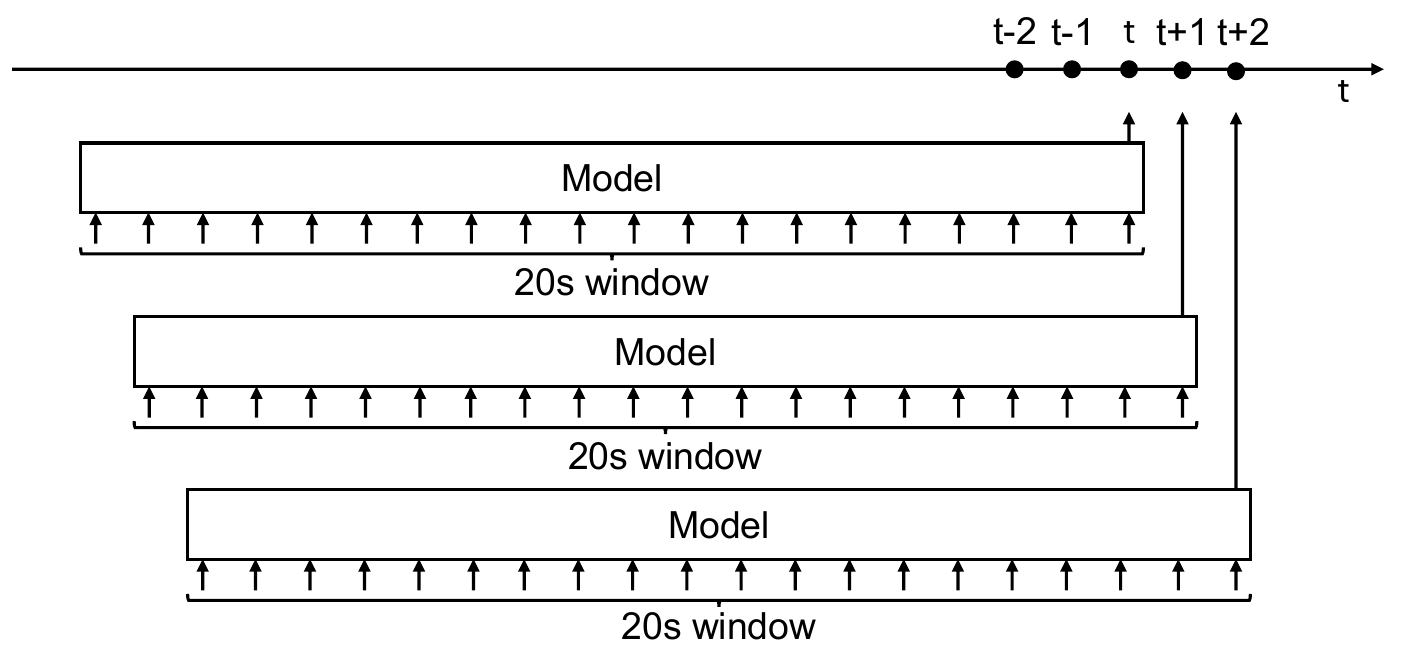}
	\caption{Time-series output using the sliding-window approach. This illustrates the case where $W=20$ and the stride is 1 second.}
    \Description{A timeline labeled t with marked time points t minus 2, t minus 1, t, t plus 1, and t plus 2. Three stacked rectangles labeled Model illustrate overlapping 20-second windows that shift forward along the timeline. The row of small arrows entering each model box from below represents the input frames within the window. Vertical arrows leaving the end of each window point to the corresponding time points, indicating outputs at successive steps.}
	\label{fig-sliding-window}
\end{figure}

In this study, we adopt a sliding-window approach that crops the entire dialogue into fixed time intervals, as shown in Figure \ref{fig-sliding-window}.
Specifically, a multimodal input of $W$ seconds, including past context, constitutes one window.
The model receives this window as input and recognizes the emotional state in the 1-second frame $t$ located at the end of the window.
The stride of the window is set to 3 seconds during training and 1 second during evaluation.

\subsection{Loss Function}\label{subsec-loss-function}

The loss function $L^{(t)}$ for frame $t$ is defined as
\begin{equation}\label{eq-loss}
L^{(t)} = w_{A}L_{A}^{(t)} + w_{V}L_{V}^{(t)} \, .
\end{equation}
Here, $w_A$ and $w_V$ are weights to adjust the contribution of each loss; both are set to 1 in all experiments.
$L_{A}^{(t)}$ and $L_{V}^{(t)}$ represent the losses for the tasks of predicting Arousal and Valence, respectively.
Since the Group Emotion labels consist of 5 classes ranging from -2 to +2 with an ordinal relationship among them, standard cross-entropy loss cannot account for such ordinal information.
Therefore, we employ the Squared Earth Mover's Distance ($\text{EMD}^2$) loss $L_{EMD^2}^{(t)}$, introduced as a loss function for deep learning by Hou et al. \cite{hou2017}.
Specifically, it is calculated using the cumulative distribution functions (CDFs) of the predicted and target distributions as
\begin{equation}\label{eq-emd}
L_{EMD^2}^{(t)} = w_t \sum_{c=-2}^{2} (CDF_{pred}^{(t)}(c) - CDF_{target}^{(t)}(c))^2
\end{equation}
where $w_t$ is the weight of frame $t$, which will be described later.
Hou et al. \cite{hou2017} demonstrated that the $\text{EMD}^2$ loss outperforms cross-entropy in ordinal classification tasks such as age estimation.
In our task, this means that predicting a neighboring class (e.g., +1 instead of +2) is penalized less than predicting a distant class (e.g., -2 instead of +2).

\subsection{Mitigation of Class Imbalance}\label{subsec-imbalance-mitigation}

As discussed in Section \ref{subsubsec-label-distribution}, the dataset used in this study has a high proportion of Class 0 in the Group Emotion labels.
To mitigate the problem of biased training data distribution, we take measures in both loss computation and data sampling.
First, using the total duration $T_c$ of each class $c$ in the training data, the total length of the training data $T$, and the number of classes $C$, we calculate the raw weight $w_c^{raw}$ for class $c$ as
\begin{equation}\label{eq-w-raw}
w_c^{raw} = 
\begin{cases} 
\frac{T}{C T_c} & (T_c > 0) \\
0 & (T_c = 0) 
\end{cases} \, .
\end{equation}
This weight is then adjusted using a smoothing factor $\alpha$ such that $w_c' = (w_c^{raw})^\alpha$, and the final weight $w_c$ for each class is defined by normalizing these values as
\begin{equation}\label{eq-w-final}
w_c = \frac{w_c'}{\frac{1}{C} \sum w_c'} \, .
\end{equation}
The weight $w_t$ for frame $t$ in $L_{EMD^2}^{(t)}$ is determined using the target class weights $w_c$.
Specifically, $w_t$ is calculated using the ground-truth label distribution $y_{c}^{(t)}$ as $w_t = \sum y_{c}^{(t)} w_c$.
Additionally, we introduce weighted sampling of the training data based on the reciprocal of the occurrence frequency of each class.

\section{Experiments}\label{sec-experiment}

We compare model types, context lengths, and input modalities, and then examine recognition performance during intervals with high Mixed values.

\subsection{Setup}\label{subsec-experiment-setup}

We evaluate the expected value of each frame-level output distribution using the Concordance Correlation Coefficient (CCC) \cite{lin1989} and Mean Absolute Error (MAE).
CCC ranges from $-1$ to $+1$ and measures both correlation and absolute agreement between predictions and ground truth.
Let $\rho$ be Pearson's correlation coefficient, $\mu_x$ and $\mu_y$ be the means of the predictions and ground truth, and $\sigma_x^2$ and $\sigma_y^2$ be their variances; CCC is defined as
\begin{equation}\label{eq-ccc}
    CCC = \frac{2 \rho \sigma_x \sigma_y}{\sigma_x^2 + \sigma_y^2 + (\mu_x - \mu_y)^2} \, .
\end{equation}
CCC is computed after concatenating all test frames.
We also report the F1 score for the positive class (Pos F1), treating classes $+1$ and $+2$ as positive and using the most probable class as the prediction.

We estimate uncertainty using a paired dialogue-level cluster bootstrap with 10,000 replicates.
In each replicate, the six test dialogues are sampled with replacement, and both conditions are evaluated on the same sample.
We report percentile 95\% confidence intervals (CIs) for the metric differences.
For the Mixed analysis, we apply the same dialogue-level resampling and compute each metric separately for HighMixed and LowMixed frames.

For CCC and MAE, the Mean baseline always outputs the mean of each training label described in Section \ref{subsubsec-label-distribution}.
For Pos F1, the Stratified baseline samples a class according to the distribution of frame-level hard labels in the training set.

In the setting names, A and V denote audio and video inputs, and the number in parentheses is the window size $W$.
We compare the Transformer and LLM under AV(20s) to examine model type, Transformer AV(5s) and AV(20s) to examine context length, and Transformer AV(20s), A(20s), and V(20s) to examine modality.
For the class-imbalance mitigation described in Section \ref{subsec-imbalance-mitigation}, we set the smoothing factor $\alpha$ to 0.5.
All models are trained for up to 10 epochs.

\subsection{Results}

\begin{table}[tb]
  \centering
  \caption{Evaluation results}
  \label{tab:results_emotion}
  \footnotesize
  \setlength{\tabcolsep}{5pt}
  \begin{tabular}{llcccccc}
    \toprule
    \multirow{2}{*}{Model} & \multirow{2}{*}{Setting} & \multicolumn{2}{c}{CCC $\uparrow$} & \multicolumn{2}{c}{MAE $\downarrow$} & \multicolumn{2}{c}{Pos F1 $\uparrow$} \\
    \cmidrule(lr){3-4} \cmidrule(lr){5-6} \cmidrule(lr){7-8}
    & & Ar & Val & Ar & Val & Ar & Val \\
    \midrule
    \multirow{4}{*}{Transformer}
      & AV(5s) & 0.787 & 0.833 & 0.124 & 0.133 & \textbf{0.632} & 0.628 \\
      & AV(20s) & \textbf{0.807} & \textbf{0.834} & \textbf{0.111} & \textbf{0.131} & 0.600 & 0.647 \\
      & A(20s) & 0.724 & 0.816 & 0.154 & 0.147 & 0.611 & 0.623 \\
      & V(20s) & 0.532 & 0.678 & 0.177 & 0.185 & 0.508 & 0.569 \\
    LLM & AV(20s) & 0.747 & 0.816 & 0.134 & 0.134 & 0.624 & \textbf{0.667} \\
    \midrule
    Mean & - & 0 & 0 & 0.163 & 0.209 & - & - \\
    Stratified & - & - & - & - & - & 0.043 & 0.067 \\
    \bottomrule
  \end{tabular}
\end{table}

Table \ref{tab:results_emotion} summarizes the results.
All trained models achieved positive CCC, compared with zero for the Mean baseline, and substantially exceeded the Stratified baseline in Pos F1.
Transformer AV(20s) achieved the highest CCC and lowest MAE for both Arousal and Valence.
Under AV(20s), the Transformer's CCC was higher than the LLM's by 0.059 for Arousal (95\% CI $[0.041, 0.087]$) and 0.018 for Valence (95\% CI $[0.005, 0.036]$), and its Arousal MAE was lower by 0.023 (95\% CI $[0.016, 0.030]$); none of these three CIs included zero.
Its Valence MAE was also lower and the LLM's Pos F1 was higher, but the CIs for these differences included zero.
Increasing the context window from 5 to 20 seconds reduced Arousal MAE by 0.012 (95\% CI $[0.004, 0.021]$); the CIs for the other metrics included zero.
For modality, the CIs excluded zero for the improvements of AV(20s) over A(20s) in CCC and MAE for both dimensions and over V(20s) in all reported metrics.
A(20s) produced better values than V(20s) for all metrics, but the CIs excluded zero only for Arousal CCC, Valence CCC, Valence MAE, and Arousal Pos F1.

\begin{figure*}[t]
  \centering
  \includegraphics[width=0.7\textwidth]{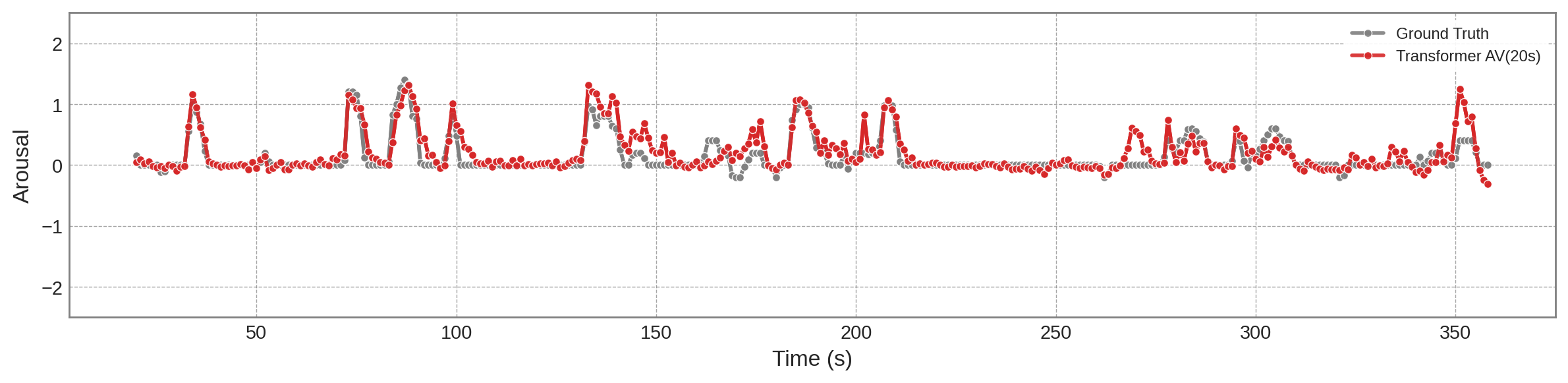}
  \caption{Output example of Arousal (Transformer AV(20s))}
  \Description{A time-series plot of Arousal versus Time (s), from about 20 to 360 seconds, with Arousal ranging from -2 to 2. Two marker-and-line traces are shown: a gray Ground Truth curve and a red Transformer AV(20s) curve. The prediction broadly follows the timing of positive peaks in the ground truth.}
  \label{fig:arousal-best-output}
\end{figure*}

\begin{figure*}[t]
  \centering
  \includegraphics[width=0.7\textwidth]{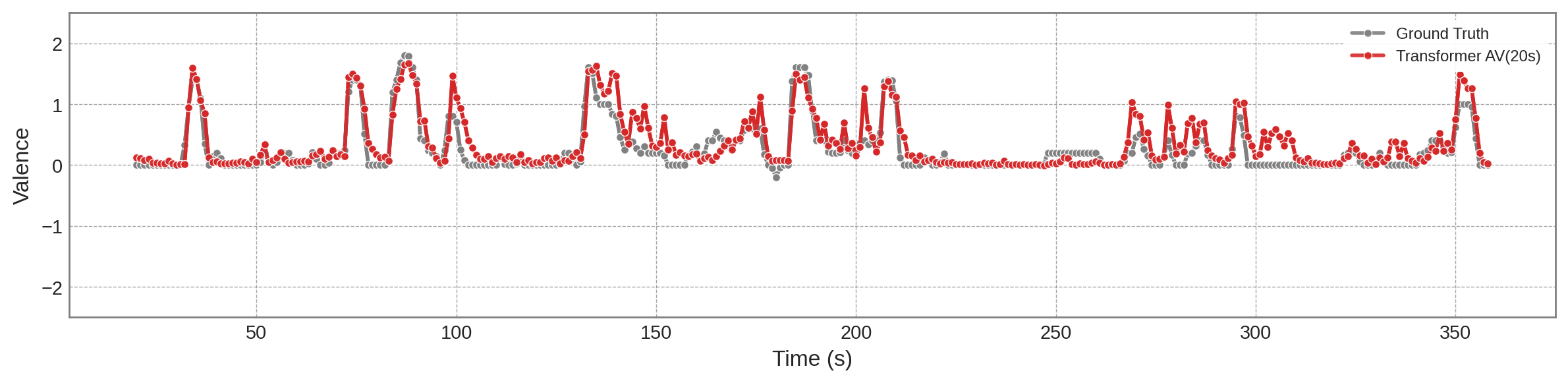}
  \caption{Output example of Valence (Transformer AV(20s))}
  \Description{A time-series plot of Valence versus Time (s), from about 20 to 360 seconds, with Valence ranging from -2 to 2. Two marker-and-line traces are shown: a gray Ground Truth curve and a red Transformer AV(20s) curve. The prediction broadly follows the timing of positive peaks in the ground truth.}
  \label{fig:valence-best-output}
\end{figure*}

Figures \ref{fig:arousal-best-output} and \ref{fig:valence-best-output} show the temporal outputs of Transformer AV(20s), which achieved the best CCC and MAE.
To examine performance during emotional divergence, we divide the test frames using the Mixed label for the corresponding dimension: HighMixed ($>0.3$) and LowMixed ($\leq0.3$).
The threshold was chosen heuristically, as higher values left too few HighMixed frames for comparison.
Table \ref{tab:results_mixed_split_transformer_av20} reports MAE and Pos F1 for each subset.
MAE was 0.177 higher in HighMixed than in LowMixed for Arousal (95\% CI $[0.126, 0.232]$) and 0.134 higher for Valence (95\% CI $[0.087, 0.188]$).
Pos F1 was also lower for HighMixed in both dimensions, but the CIs for these differences included zero.

\begin{table}[tb]
  \centering
  \caption{Comparison of model performance by Mixed value (Transformer AV(20s), threshold = 0.3)}
  \label{tab:results_mixed_split_transformer_av20}
  \footnotesize
  \setlength{\tabcolsep}{8.5pt}
  \begin{tabular}{llccc}
    \toprule
    Dimension & Condition & Samples & MAE $\downarrow$ & Pos F1 $\uparrow$ \\
    \midrule
    \multirow{2}{*}{Arousal} & LowMixed ($\leq 0.3$) & 2002 & 0.101 & 0.654 \\
            & HighMixed ($> 0.3$) & 124 & 0.278 & 0.407 \\
    \midrule
    \multirow{2}{*}{Valence} & LowMixed ($\leq 0.3$) & 2042 & 0.126 & 0.658 \\
            & HighMixed ($> 0.3$) & 84 & 0.259 & 0.545 \\
    \bottomrule
  \end{tabular}
\end{table}

\section{Discussion}

Under the same AV(20s) setting, the Transformer achieved higher CCC for both dimensions and lower Arousal MAE than the LLM, with dialogue-level bootstrap CIs excluding zero for these differences.
This suggests that the temporal Transformer was more effective for continuous estimation in the current setting.
By contrast, the CIs for the LLM's higher Pos F1 values included zero, so the results do not show an advantage in detecting positive states.
This comparison does not establish a general limitation of LLMs: the current multimodal interface and amount of training data may not fully exploit their capacity, while language-based reasoning and explanation remain promising uses of LLMs for this task.

Among the context-length comparisons, only the CI for Arousal MAE excluded zero in favor of the 20-second context.
Thus, the present results do not provide consistent evidence that a longer context improves recognition.
For the audio-visual model, the CIs for CCC and MAE improvements over both unimodal models excluded zero.
Audio also outperformed video on several metrics, suggesting that audio was the stronger unimodal signal in this setting while video provided complementary information when combined with audio.
The associations of Group Emotion with laughter and dialogue activity described in Section \ref{subsec-label-dialog-correlation} may partly explain the importance of audio.

MAE was higher during HighMixed intervals for both dimensions, with the dialogue-level bootstrap CIs excluding zero.
Emotional divergence is therefore associated with larger errors when Group Emotion is represented by a single value.
One possible explanation is that the model responds strongly to salient emotions expressed by only some participants, even when those emotions do not determine the annotated group-level value.
A second explanation is that the ground-truth values are less reliable in these intervals, since annotators had to assign a single value even in segments they labeled as Mixed.
The CIs for the decrease in Pos F1 included zero, and the number of HighMixed frames was limited.
Nevertheless, the MAE results motivate models that explicitly represent the coexistence of multiple participant states.

Several limitations qualify these conclusions.
Although complete three-person groups were held out from training, some individuals appeared in multiple groups; the evaluation therefore measures generalization to unseen group configurations and interactions rather than unseen individuals.
Because the six test dialogues come from only two groups, dialogues that share participants are not independent, and the reported dialogue-level bootstrap intervals are likely narrower than group-level intervals would be.
Each condition was also trained once with a single data split, so variation due to the choice of test groups and to parameter initialization was not assessed.
Finally, the annotations lack participant-level emotion labels: Mixed indicates when annotators judged a single group-level value to be inadequate, but not how individual states relate to the group-level label.
Future evaluations should therefore use person-independent splits and multiple training runs, and future annotations should cover both participant- and group-level states.

\section{Conclusion}

This study addressed the continuous recognition of Group Emotion (Arousal and Valence) in multi-party dialogue.
We continuously annotated a multi-party dialogue corpus at a 1-second resolution, developed and evaluated a multimodal temporal framework on this dataset, and analyzed emotional divergence using Mixed labels.
The temporal Transformer outperformed distribution-based baselines and the LLM-based model in CCC for both dimensions and in Arousal MAE.
The 20-second context provided only a limited improvement over the 5-second context, while combining audio and video improved CCC and MAE over either modality alone, with audio providing the stronger unimodal signal.
Recognition errors were larger during Mixed intervals, with dialogue-level bootstrap CIs excluding zero for these differences, consistent with the difficulty of representing divergent participant emotions using a single group-level value.

These findings, together with the limitations discussed above, suggest three directions for future work.
First, annotations of both participant- and group-level states would let models represent individual emotions explicitly and relate them to the group-level state, addressing the errors observed during Mixed intervals.
Second, identifying which visual cues provide reliable complementary information to audio remains an open problem.
Third, the sliding-window inference adopted here is computationally expensive, and real-time operation will require streaming architectures that maintain historical context efficiently.

\section*{Safe and Responsible Innovation Statement}

This study proposes continuous Group Emotion recognition using multimodal temporal models to support natural and smooth Human-Computer Interaction (HCI).
A primary concern regarding potential misuse is the application of this technology for group surveillance.
The dialogue dataset used in this study was strictly managed, with informed consent obtained from all participants in advance.
Furthermore, as the participants in the dialogue data belong to a specific cultural group, there is a possibility that the model contains cultural biases.
Therefore, careful consideration of diverse cultural backgrounds is necessary when deploying this system in real-world society.

\begin{acks}
  This work was supported by JST Moonshot R\&D (JPMJPS2011).
\end{acks}

\bibliographystyle{ACM-Reference-Format}
\bibliography{references}

\end{document}